\documentclass[conference]{IEEEtran}
\IEEEoverridecommandlockouts
\usepackage{amsmath}
\usepackage{amssymb}
\usepackage{booktabs}
\usepackage{graphicx}
\usepackage{cite}
\usepackage{amsmath,amssymb,amsfonts}
\usepackage{algorithmic}
\usepackage{graphicx}
\usepackage{textcomp}
\usepackage{xcolor}
\def\BibTeX{{\rm B\kern-.05em{\sc i\kern-.025em b}\kern-.08em
    T\kern-.1667em\lower.7ex\hbox{E}\kern-.125emX}}
\begin{document}

\title{Secure Multimodal Federated Optimization via Hierarchical DNA-Based Key Management\\
%{\footnotesize \textsuperscript{*}Note: Sub-titles are not captured for https://ieeexplore.ieee.org  and should not be used}
\thanks{This work is funded by the International Collaboration Fund for Creative Research of National Science Foundation of China (No. W2441019), the China Postdoctoral Science Foundation (No. 2026M791699), and the Young Scientists Fund of the National Natural Science Foundation of China (No. 6260070105).}
}

\author{\IEEEauthorblockN{1\textsuperscript{st} Yuping Yan}
\IEEEauthorblockA{\textit{School of Engineering} \\
\textit{Westlake University}\\
Hangzhou, China \\
yanyuping@westlake.edu.cn}
\and
\IEEEauthorblockN{2\textsuperscript{nd} Yaochu Jin\textsuperscript{*}\thanks{*Corresponding author: jinyaochu@westlake.edu.cn}}
\IEEEauthorblockA{\textit{School of Engineering} \\
\textit{Westlake University}\\
Hangzhou, China \\
jinyaochu@westlake.edu.cn}
%\and
%\IEEEauthorblockN{3\textsuperscript{rd} Given Name Surname}
%\IEEEauthorblockA{\textit{dept. name of organization (of Aff.)} \\
%\textit{name of organization (of Aff.)}\\
%City, Country \\
%email address or ORCID}
%\and
%\IEEEauthorblockN{4\textsuperscript{th} Given Name Surname}
%\IEEEauthorblockA{\textit{dept. name of organization (of Aff.)} \\
%\textit{name of organization (of Aff.)}\\
%City, Country \\
%email address or ORCID}
%\and
%\IEEEauthorblockN{5\textsuperscript{th} Given Name Surname}
%\IEEEauthorblockA{\textit{dept. name of organization (of Aff.)} \\
%\textit{name of organization (of Aff.)}\\
%City, Country \\
%email address or ORCID}
%\and
%\IEEEauthorblockN{6\textsuperscript{th} Given Name Surname}
%\IEEEauthorblockA{\textit{dept. name of organization (of Aff.)} \\
%\textit{name of organization (of Aff.)}\\
%City, Country \\
%email address or ORCID}
}

\maketitle

\begin{abstract}
Federated optimization over heterogeneous multimodal data enables distributed institutions to collaboratively optimize a shared model using medical images, genomic profiles, clinical records, and physiological signals without centralizing raw data. However, protecting modality-specific data throughout the optimization lifecycle remains insufficiently explored, particularly when heterogeneous modalities require separate encryption keys and selective local access. This paper proposes a secure multimodal federated optimization framework that separates local data protection from federated-update protection. A trusted key-management authority securely provisions one client key through Hybrid Public Key Encryption (HPKE), from which each client locally derives modality-specific symmetric keys using the domain-separated HMAC-based Extract-and-Expand Key Derivation Function (HKDF). Heterogeneous modalities are mapped into a unified DNA-domain representation and processed through a DNA-origami-inspired structural transformation before authenticated encryption with AES-GCM. During local optimization, only the required modalities are decrypted within an authorized client environment. Individual model updates are independently protected through secure aggregation, allowing the coordinator to observe only aggregated updates rather than each client's contribution. This separation addresses complementary protection requirements for locally stored multimodal data and information exchanged during collaborative optimization. The proposed hierarchy requires only one persistent secret per client, avoiding independent provisioning of every modality key. Compromise of an operational modality key is confined to its associated modality, provided that the client key remains secure. We further provide compositional security and complexity analyses to characterize its security guarantees, key-management costs, and overall computational overhead. Together, these components establish a structured approach to modality-specific protection and secure collaborative optimization, integrating hierarchical key management, authenticated local encryption, and aggregation-level update protection within a unified framework.
\end{abstract}


\section{Introduction}
\label{sec:introduction}

Biomedical analysis increasingly integrates heterogeneous and complementary data modalities, including medical images, genomic profiles, electronic health records, and physiological signals. Jointly exploiting these modalities can improve disease prediction, clinical decision support, and automated biomedical analysis~\cite{acosta2022multimodal}. However, centralizing such sensitive data is often impractical because of privacy concerns, institutional boundaries, and regulatory constraints~\cite{voigt2017eu}. Federated optimization offers a promising alternative by allowing multiple institutions to collaboratively optimize a global model while retaining their raw data locally~\cite{liu2024secure}.

Multimodal federated optimization nevertheless introduces security challenges beyond conventional non-independent and identically distributed (non-IID) statistical heterogeneity \cite{yan2024dp}. Different institutions may possess different modality subsets, while individual modalities vary in representation, sensitivity, and protection requirements. Existing privacy-preserving federated methods commonly focus on protecting communicated model updates or apply a uniform protection policy to all local information~\cite{yan2024fedlabx}. Consequently, they do not explicitly isolate the security domains of different modalities or jointly address local multimodal-data protection and modality-aware key management.

Key management creates an additional security--efficiency trade-off. Using one client-wide key minimizes storage and provisioning costs, but its compromise may expose all modality payloads protected under that key~\cite{barker2018recommendation}. Conversely, maintaining an independent long-term key for every modality provides finer-grained isolation but incurs substantial generation, provisioning, storage, rotation, and revocation overhead. A practical solution should therefore contain an operational-key compromise within one modality while requiring only limited persistent key material.

To address these challenges, we propose a hierarchical key-managed framework for secure multimodal federated optimization. A trusted key-management authority provisions one client key through Hybrid Public Key Encryption (HPKE), from which the client derives modality-specific operational keys using the HMAC-based Extract-and-Expand Key Derivation Function (HKDF). Heterogeneous modalities are mapped into a common DNA-domain representation, structurally transformed, and protected using the Advanced Encryption Standard in Galois/Counter Mode (AES-GCM). Individual optimization updates are protected separately through secure aggregation (SecAgg), ensuring that the coordinator learns only their aggregate. The DNA-domain transformation provides a unified representation interface, whereas standard cryptographic confidentiality and integrity are supplied by AES-GCM. The main contributions are summarized as follows:

\begin{itemize}
    \item We formulate secure multimodal federated optimization under both statistical and modality heterogeneity, and distinguish the local multimodal-data boundary from the federated-update boundary.

    \item We develop a lightweight hierarchical protection framework that integrates HPKE-based client-key provisioning, HKDF-based modality-key derivation, DNA-domain representation, and AES-GCM authenticated encryption processes. Each client stores only one persistent key, while compromise of an operational modality key is confined to its associated modality.

   \item We provide a compositional security analysis under standard assumptions on the underlying cryptographic primitives, together with an overhead analysis of key management. The analysis demonstrates that the proposed hierarchy achieves efficient modality-aware protection with only one persistent client key and linear modality-key derivation cost.
\end{itemize}
\section{Related Work}
\label{sec:related_work}

\subsection{Multimodal Optimization and Federated Learning}

Recent optimization methods exploit structured and multimodal representations to improve search and problem modeling. Reijnen \emph{et al.} encode evolutionary search states as graphs for dynamic algorithm configuration~\cite{reijnen2025graph}, while Chen \emph{et al.} fuse graph and image representations for neural multi-objective combinatorial optimization~\cite{chen2025neural}. In distributed settings, multimodal federated learning (MMFL) mainly addresses modality-aware representation learning, heterogeneous fusion, and missing modalities. Representative studies include unified modality-specific learning~\cite{xiong2022unified}, correlation-adaptive aggregation across modality-incongruent clients~\cite{chen2022fedmsplit}, and the FedMultimodal benchmark~\cite{feng2023fedmultimodal}. Recent methods further improve communication efficiency through modality and client selection~\cite{ma2026fedmc}, or handle missing modalities through prototype transfer and representation alignment~\cite{le2025cross}. However, secure optimization over heterogeneous modalities under distributed data ownership remains insufficiently explored.

\subsection{Privacy Protection and Key Management}

Recent MMFL studies have begun to address privacy. TriCon-SF employs split--shuffle operations to protect distributed aggregation~\cite{yan2025tricon}, while Sec-MMFL estimates modality-dependent leakage and adaptively adjusts protection strength~\cite{xiao2026enhancing}. DNA-origami cryptography provides an alternative structural protection mechanism by encoding messages into DNA patterns recoverable through key-controlled folding~\cite{zhang2019dna}. Nevertheless, these approaches address update protection, privacy allocation, or DNA-based encoding separately. A unified framework for modality-aware key provisioning, local multimodal-data protection, compromise containment, and federated-update privacy remains underexplored.

\section{Preliminaries}
\label{sec:preliminaries}

\subsection{DNA-Origami-Inspired Structural Encoding}
\label{subsec:dna_primitive}

DNA origami folds a long scaffold strand into an addressable
nanostructure using a set of short staple strands. DNA origami cryptography encodes a message as a permuted Braille-like spot pattern, maps the pattern to message strands on the scaffold, and reveals it only after reconstruction using the corresponding folding configuration
\cite{zhang2019dna}.

We abstract this physical process as an invertible structural
transformation. For modality $m$, the encoder first maps data
$\mathbf{x}_i^{(m)}$ to a DNA sequence
\begin{equation}
S_i^{(m)}
=
\phi_m(\mathbf{x}_i^{(m)})
\in \Sigma_{\rm DNA}^{L_m},
\qquad
\Sigma_{\rm DNA}=\{A,C,G,T\}.
\end{equation}
A folding configuration $\theta_{i,e}^{(m)}$ then determines the arrangement and permutation of the sequence:
\begin{equation}
F_i^{(m)}
=
\mathsf{Fold}_{\theta_{i,e}^{(m)}}(S_i^{(m)}),
\qquad
S_i^{(m)}
=
\mathsf{Unfold}_{\theta_{i,e}^{(m)}}(F_i^{(m)}).
\label{eq:dna_primitive}
\end{equation}
The transformation provides a common structural representation for heterogeneous modalities. It is not treated as the sole source of cryptographic confidentiality, which is instead provided by authenticated symmetric encryption.

\subsection{Cryptographic Building Blocks}
\label{subsec:crypto_primitives}

\textbf{HPKE.}
Hybrid Public Key Encryption (HPKE) combines public-key encapsulation, key derivation, and authenticated encryption. It is used only to provision the client key:
\begin{equation}
\widehat{K}_i
=\mathsf{HPKE.Seal}(pk_i,K_i),
\qquad
K_i=\mathsf{HPKE.Open}(sk_i,\widehat{K}_i).
\end{equation}

\textbf{HKDF.}
The HMAC-based Extract-and-Expand Key Derivation Function (HKDF) derives modality-specific operational material from $K_i$. We define
\begin{equation}
\mathsf{OKM}_{i,e}^{(m)}
=
\mathsf{HKDF}
\left(
K_i,
\texttt{modality}\Vert ID_i\Vert ID_m\Vert e
\right),
\end{equation}
and parse the output as
\begin{equation}
\mathsf{OKM}_{i,e}^{(m)}
=
K_{i,e}^{(m)}
\Vert
\theta_{i,e}^{(m)},
\label{eq:operational_material}
\end{equation}
where $K_{i,e}^{(m)}$ is the modality encryption key and
$\theta_{i,e}^{(m)}$ is the seed used to generate its folding
configuration.

\textbf{Authenticated encryption.}
The folded representation is protected using AES-GCM:
\begin{equation}
(C_i^{(m)},\tau_i^{(m)})
=
\mathsf{AEAD.Enc}_{K_{i,e}^{(m)}}
\left(
n_i^{(m)},F_i^{(m)},A_i^{(m)}
\right),
\end{equation}
where $n_i^{(m)}$ is a unique nonce, $A_i^{(m)}$ is associated data, and $\tau_i^{(m)}$ is the authentication tag.

\textbf{Secure aggregation.}
For local updates $\{\Delta\mathbf{w}_i^t\}_{i\in\mathcal{S}_t}$,
secure aggregation reveals only
\begin{equation}
\mathsf{SecAgg}
\left(
\{\alpha_i\Delta\mathbf{w}_i^t\}_{i\in\mathcal{S}_t}
\right)
=
\sum_{i\in\mathcal{S}_t}
\alpha_i\Delta\mathbf{w}_i^t,
\end{equation}
without disclosing an individual client update to the coordinator.

\section{Problem Formulation and Threat Model}
\label{sec:problem}

\subsection{Multimodal Federated Optimization}
\label{subsec:problem_formulation}

We consider a federated optimization system consisting of a coordinator and $N$ clients $\mathcal{C}=\{C_1,\ldots,C_N\}$. Let $\mathcal{M}=\{1,\ldots,M\}$ denote the complete modality set, including, for example, medical images, genomic profiles, clinical records, and physiological signals. Client $C_i$ holds only a subset $\mathcal{M}_i\subseteq\mathcal{M}$, with local data
\begin{equation}
\mathcal{D}_i
=
\left\{
\mathcal{D}_i^{(m)}
\mid
m\in\mathcal{M}_i
\right\}.
\label{eq:local_multimodal_data}
\end{equation}
Since $\mathcal{M}_i$ may differ across clients, the system exhibits \emph{modality heterogeneity} in addition to conventional non-IID statistical heterogeneity.

The clients collaboratively optimize a global parameter vector $\mathbf{w}$ without sharing their raw multimodal data:
\begin{equation}
\min_{\mathbf{w}}
F(\mathbf{w})
=
\sum_{i=1}^{N}
p_iF_i(\mathbf{w};\mathcal{D}_i),
\qquad
\sum_{i=1}^{N}p_i=1,
\label{eq:global_objective}
\end{equation}
where $F_i$ is the local objective and $p_i$ is the contribution weight of client $C_i$. At round $t$, a selected client performs a generic local optimization procedure
\begin{equation}
\mathbf{w}_i^{t+1}
=
\mathcal{O}_i
\left(
\mathbf{w}^{t};
\mathcal{D}_i
\right),
\qquad
\Delta\mathbf{w}_i^{t}
=
\mathbf{w}_i^{t+1}-\mathbf{w}^{t},
\label{eq:local_optimization}
\end{equation}
where $\mathcal{O}_i$ may be gradient-based, evolutionary, or otherwise data-driven. The coordinator updates the global state using the aggregated local updates.

\subsection{Protection Objectives}
\label{subsec:security_objectives}

The framework protects two distinct security boundaries. At the \emph{local data boundary}, raw multimodal data and sensitive representations are protected using modality-specific symmetric keys. At the \emph{federation boundary}, individual optimization updates are hidden through secure aggregation.

A trusted key-management authority maintains a root key $K_0$ and provisions one client key $K_i$ to each client through authenticated public-key encryption. Client $C_i$ then locally derives its modality keys:
\begin{equation}
K_0
\rightarrow
K_i
\rightarrow
\left\{
K_{i,e}^{(m)}
\right\}_{m\in\mathcal{M}_i},
\label{eq:key_hierarchy_problem}
\end{equation}
where $e$ denotes the key epoch. Only $K_{i,e}^{(m)}$ is used for symmetric encryption and decryption of modality $m$.

The desired key-separation properties are
\begin{equation}
K_i\nRightarrow K_j,\quad i\neq j,
\qquad
K_{i,e}^{(m)}
\nRightarrow
K_{i,e}^{(m')},\quad m\neq m',
\label{eq:key_isolation_problem}
\end{equation}
where $\nRightarrow$ denotes computational infeasibility. Thus, exposure of one modality key should not reveal its parent, sibling modalities, or keys belonging to another client.

For update privacy, the coordinator should learn only the aggregate
\begin{equation}
\Delta\mathbf{w}^{t}
=
\sum_{i\in\mathcal{S}_t}
\alpha_i\Delta\mathbf{w}_i^{t},
\label{eq:aggregate_only}
\end{equation}
rather than any individual $\Delta\mathbf{w}_i^{t}$.

\subsection{Threat Model}
\label{subsec:threat}

We consider an honest-but-curious coordinator that follows the federated optimization protocol but may infer private information from received updates and metadata. It has no access to raw multimodal data or client-side keys and does not collude with the trusted key-management authority. We also consider a passive external adversary that can intercept protected communications. Client keys are provisioned through authenticated HPKE, while multimodal payloads are protected by AES-GCM; forged or modified ciphertexts are rejected upon authentication failure.

We further allow partial key exposure through credential leakage or unauthorized local access. Compromise of a modality key $K_{i,e}^{(m)}$ should remain confined to modality $m$ and reveal neither sibling modality keys nor keys of other clients. Compromise of the parent key $K_i$ may expose its descendant modality keys, but not those of any $C_j$, $j\neq i$. We assume that $K_0$, uncompromised client private keys, and the underlying HPKE, HKDF, AES-GCM, and secure-aggregation primitives remain secure. Root-key compromise, coordinator--authority collusion, Byzantine behavior, model poisoning, denial-of-service, and complete endpoint takeover are outside the scope of this work.

\begin{figure*}
    \centering
    \includegraphics[width=\linewidth]{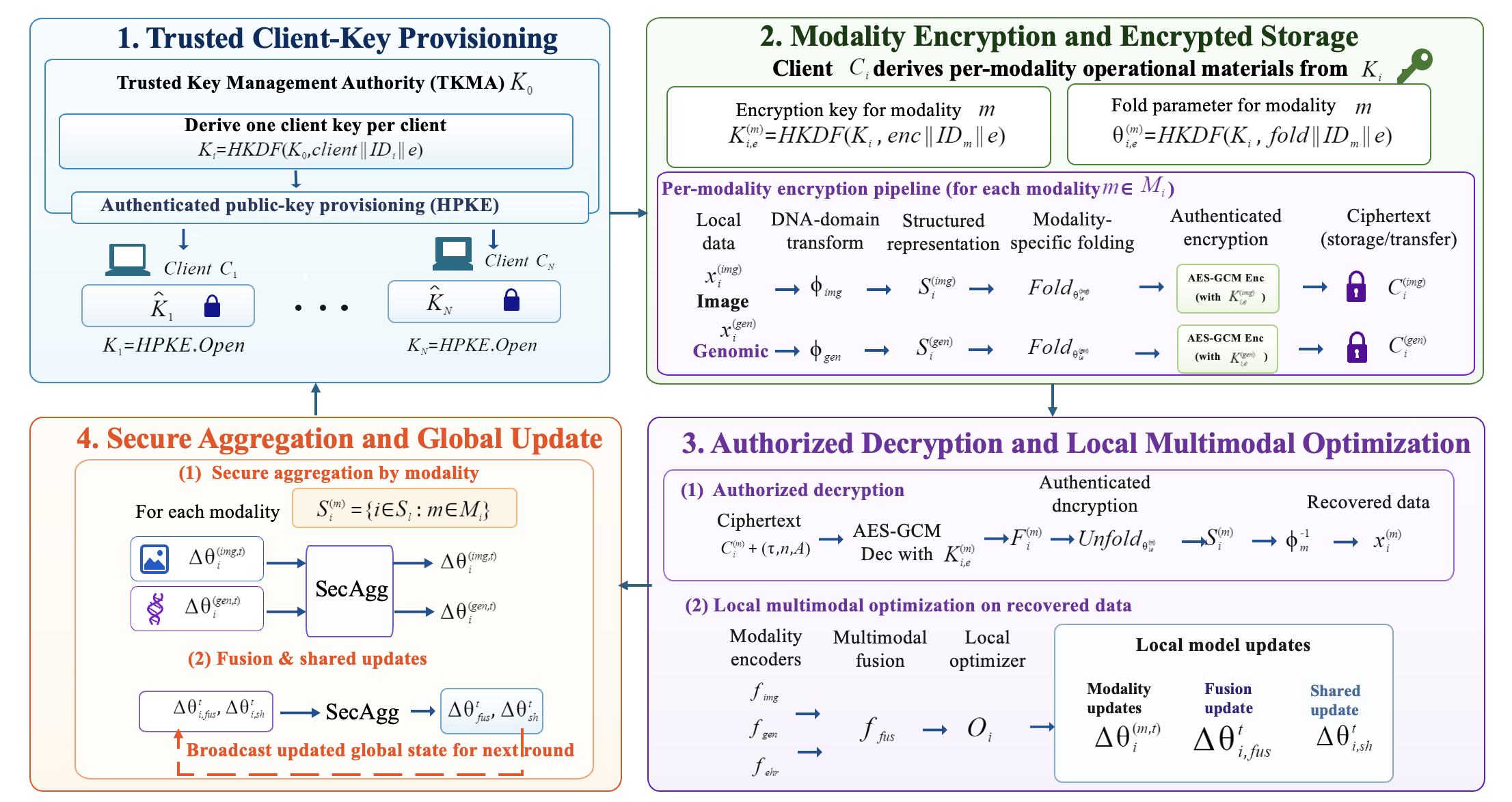}
    \caption{Overview of Hierarchical DNA-Based Key Management in Multi-modal Federated Optimization}
    \label{fig:framework}
\end{figure*}

\section{Methodology}
\label{sec:framework}

To address the multimodal protection and key-management challenges formulated in Section~\ref{sec:problem}, we propose a secure multimodal federated optimization framework that integrates hierarchical key management, DNA-based multimodal protection, and secure model-update aggregation. As illustrated in Fig.~\ref{fig:framework}, the framework separates three functional planes: 1) a \emph{key-management plane} for client-key provisioning and modality-key derivation; 2) a \emph{data-protection plane} for encrypting heterogeneous multimodal data and representations; and 3) an \emph{optimization plane} for securely aggregating local optimization updates.

The key hierarchy is simplified as
\begin{equation}
K_0
\rightarrow
K_i
\rightarrow
\left\{
K_i^{(m)}
\right\}_{m\in\mathcal{M}_i},
\label{eq:simplified_hierarchy}
\end{equation}
where $K_0$, $K_i$, and $K_i^{(m)}$ denote the system root key, client key, and modality-specific operational key, respectively. Only $K_i^{(m)}$ is used to encrypt multimodal information. Model updates are protected separately through secure aggregation and are never encrypted using modality keys.

\subsection{Client-Key Provisioning}
\label{subsec:key_provisioning}

A trusted key-management authority, which is independent of the federated optimization coordinator, maintains the root secret $K_0$. For each registered client $C_i$, it derives a client-specific key as
\begin{equation}
K_i
=
\operatorname{HKDF}
\left(
K_0,
\texttt{client}\Vert ID_i\Vert e
\right),
\label{eq:client_key_new}
\end{equation}
where $ID_i$ is the client identifier, $e$ denotes the current key epoch, and $\Vert$ represents concatenation. The explicit \texttt{client} label provides domain separation between different key purposes.

Each client owns a public--private key pair $(pk_i,sk_i)$. The key-management authority securely provisions $K_i$ using hybrid public-key encryption:
\begin{equation}
\widehat{K}_i
=
\operatorname{HPKE.Seal}
\left(
pk_i,K_i;A_i^{\rm key}
\right),
\label{eq:hpke_seal}
\end{equation}
where $A_i^{\rm key}$ contains authenticated provisioning metadata such as $ID_i$ and the key epoch. Client $C_i$ recovers the key using
\begin{equation}
K_i
=
\operatorname{HPKE.Open}
\left(
sk_i,\widehat{K}_i;A_i^{\rm key}
\right).
\label{eq:hpke_open}
\end{equation}

Thus, the provisioning stage uses asymmetric cryptography: the authority encrypts with $pk_i$, whereas the client decrypts with the distinct private key $sk_i$. The coordinator neither participates in this process nor receives the client key.

\subsection{Local Modality-Key Derivation}
\label{subsec:modality_key_derivation}

After receiving $K_i$, client $C_i$ locally derives one symmetric key for each available modality:
\begin{equation}
K_{i,e}^{(m)}
=
\operatorname{HKDF}
\left(
K_i,
\texttt{modality}
\Vert ID_m
\Vert e
\right),
\qquad
m\in\mathcal{M}_i,
\label{eq:modality_key_new}
\end{equation}
where $ID_m$ identifies the modality, such as image, genomic, clinical, or physiological data.

Distinct modality identifiers produce cryptographically separated operational keys:
\begin{equation}
K_{i,e}^{(m)}
\neq
K_{i,e}^{(m')},
\qquad
m\neq m'.
\end{equation}
Consequently, compromise of one modality key does not directly reveal the keys protecting the other modalities.

Only $K_i$ needs to be stored persistently by the client. Modality keys can be derived on demand, cached within one key epoch, and erased after use. Therefore, the client obtains modality-level compromise isolation without persistently maintaining a large collection of independent keys.

\subsection{DNA-Based Multimodal Protection}
\label{subsec:dna_protection}

The data-protection plane provides a unified interface for heterogeneous modalities. Let
\begin{equation}
\Sigma_{\rm DNA}=\{A,C,G,T\}
\end{equation}
denote the DNA alphabet. For each modality $m$, a modality-specific encoder maps the original data into a DNA-domain representation:
\begin{equation}
S_i^{(m)}
=
\phi_m
\left(
\mathbf{x}_i^{(m)}
\right)
\in
\Sigma_{\rm DNA}^{L_m}.
\label{eq:dna_encode_new}
\end{equation}

The encoder $\phi_m$ is adapted to the structure of modality $m$. Images and numerical records can be quantized, serialized, and mapped to DNA bases, while sequence-type modalities can be transformed using their corresponding symbol-to-base mappings. This converts heterogeneous input spaces into a common DNA representation without requiring identical modality structures.

A DNA-origami-inspired transformation is then applied:
\begin{equation}
F_i^{(m)}
=
\operatorname{Fold}
\left(
S_i^{(m)};
\theta_i^{(m)}
\right),
\label{eq:dna_fold}
\end{equation}
where $\theta_i^{(m)}$ determines the modality-specific arrangement, permutation, and folding configuration. This operation provides a structured representation transformation but does not replace the standard cryptographic encryption primitive.

The transformed representation is protected using authenticated symmetric encryption, instantiated by AES-GCM:
\begin{equation}
\left(
C_i^{(m)},\tau_i^{(m)}
\right)
=
\operatorname{AEAD.Enc}_{K_{i,e}^{(m)}}
\left(
n_i^{(m)},
F_i^{(m)},
A_i^{(m)}
\right),
\label{eq:aead_encrypt}
\end{equation}
where $C_i^{(m)}$ is the ciphertext, $\tau_i^{(m)}$ is the authentication tag, $n_i^{(m)}$ is a unique nonce, and $A_i^{(m)}$ denotes associated data.

The associated data binds the ciphertext to its security context:
\begin{equation}
A_i^{(m)}
=
ID_i
\Vert ID_m
\Vert e
\Vert t
\Vert \texttt{type},
\label{eq:aad}
\end{equation}
where $t$ is the federated optimization round and \texttt{type} distinguishes raw data from an intermediate representation. Therefore, different data objects and optimization rounds can safely use the same modality key, provided that their nonces are unique.

An authorized client verifies and decrypts the protected representation as
\begin{equation}
F_i^{(m)}
=
\operatorname{AEAD.Dec}_{K_{i,e}^{(m)}}
\left(
n_i^{(m)},
C_i^{(m)},
A_i^{(m)},
\tau_i^{(m)}
\right).
\label{eq:aead_decrypt}
\end{equation}
If authentication fails, the ciphertext is rejected. Otherwise, the original modality is recovered through
\begin{equation}
\mathbf{x}_i^{(m)}
=
\phi_m^{-1}
\left(
\operatorname{Unfold}
\left(
F_i^{(m)};
\theta_i^{(m)}
\right)
\right).
\label{eq:dna_recovery}
\end{equation}

Because AES-GCM is symmetric, encryption and decryption use the same modality key:
\begin{equation}
K_{\rm enc}
=
K_{\rm dec}
=
K_{i,e}^{(m)}.
\end{equation}
The complete modality-protection pipeline is therefore
\begin{equation}
\mathbf{x}_i^{(m)}
\xrightarrow{\phi_m}
S_i^{(m)}
\xrightarrow{\operatorname{Fold}}
F_i^{(m)}
\xrightarrow[\;K_{i,e}^{(m)}\;]{\operatorname{AEAD.Enc}}
C_i^{(m)},
\label{eq:forward_pipeline}
\end{equation}
with authorized recovery performed in the reverse direction.

\subsection{Secure Multimodal Federated Optimization}
\label{subsec:secure_fedopt}

At communication round $t$, the coordinator distributes the current global optimization state $\mathbf{w}^{t}$ to the selected client set $\mathcal{S}_t$. Each client decrypts only the locally required modalities and performs optimization within its trusted local environment.

For client $C_i$, modality-specific representations are obtained as
\begin{equation}
\mathbf{z}_i^{(m)}
=
f_m
\left(
\mathbf{x}_i^{(m)};
\mathbf{w}^{t}
\right),
\qquad
m\in\mathcal{M}_i,
\end{equation}
and integrated using
\begin{equation}
\mathbf{z}_i
=
\mathcal{F}
\left(
\{
\mathbf{z}_i^{(m)}
\}_{m\in\mathcal{M}_i}
\right).
\label{eq:multimodal_fusion_new}
\end{equation}

The local optimization procedure is written generally as
\begin{equation}
\mathbf{w}_i^{t+1}
=
\mathcal{O}_i
\left(
\mathbf{w}^{t};
\mathcal{D}_i,
\mathbf{z}_i
\right),
\label{eq:local_optimizer}
\end{equation}
where $\mathcal{O}_i$ may represent a gradient-based, evolutionary, or other data-driven local optimizer. The resulting local update is
\begin{equation}
\Delta\mathbf{w}_i^{t}
=
\mathbf{w}_i^{t+1}
-
\mathbf{w}^{t}.
\label{eq:local_update_new}
\end{equation}

Modality keys are not used to encrypt $\Delta\mathbf{w}_i^{t}$ because providing such keys to the coordinator would violate modality-data confidentiality. Instead, participating clients execute a secure aggregation protocol:
\begin{equation}
\Delta\mathbf{w}^{t}
=
\operatorname{SecAgg}
\left(
\{
\alpha_i
\Delta\mathbf{w}_i^{t}
\}_{i\in\mathcal{S}_t}
\right)
=
\sum_{i\in\mathcal{S}_t}
\alpha_i
\Delta\mathbf{w}_i^{t},
\label{eq:secure_aggregation_new}
\end{equation}
where $\alpha_i$ is the contribution weight of client $C_i$ and
$\sum_{i\in\mathcal{S}_t}\alpha_i=1$.

The coordinator observes only the aggregated result and updates the global optimization state:
\begin{equation}
\mathbf{w}^{t+1}
=
\mathbf{w}^{t}
+
\Delta\mathbf{w}^{t}.
\label{eq:global_update_new}
\end{equation}

\section{Security and Complexity Analysis}
\label{sec:analysis}

We analyze the security and overhead of the proposed framework under the following assumptions: the root key $K_0$ and uncompromised parent keys remain secret; HKDF provides pseudorandom, domain-separated key derivation; HPKE and AES-GCM provide their standard security guarantees; and the adopted secure aggregation protocol hides individual client updates under its required participation assumptions.

\subsection{Security Proof}
\label{subsec:security_proof}

Let $\operatorname{Adv}_{\Pi}^{\rm sec}(\mathcal A)$ denote the
advantage of a probabilistic polynomial-time adversary in recovering
or predicting an uncompromised key, distinguishing protected modality
data, producing an accepted ciphertext forgery, or learning more than
the permitted aggregate from secure aggregation.

\textbf{Theorem 1 (Compositional Security).}
Assume authenticated client public keys, an uncompromised root key
$K_0$, domain-separated HKDF, confidential HPKE, nonce-respecting
AEAD confidentiality and integrity, and a privacy-preserving secure
aggregation protocol. Then there exist PPT reductions
$\mathcal B_1,\ldots,\mathcal B_5$ such that
\begin{align} 
\operatorname{Adv}_{\Pi}^{\rm sec}(\mathcal{A}) \leq\; \operatorname{Adv}_{\rm HPKE}^{\rm conf}(\mathcal{B}_1) +q_K\operatorname{Adv}_{\rm HKDF}^{\rm prf}(\mathcal{B}_2) \nonumber\\ +q_E\!\left[ \operatorname{Adv}_{\rm AEAD}^{\rm conf}(\mathcal{B}_3) +\operatorname{Adv}_{\rm AEAD}^{\rm int}(\mathcal{B}_4) \right]+\operatorname{Adv}_{\rm SA}^{\rm priv}(\mathcal{B}_5), 
\label{eq:security_bound} 
\end{align}
where $q_P$, $q_K$, $q_E$, and $q_F$ denote the numbers of key
provisionings, derived keying materials, protected payloads, and
verification attempts, respectively.

\emph{Proof sketch.}
Let $H_0$ be the real protocol. In $H_1$, client keys and uncompromised
modality keying materials are replaced, one at a time, by independent
random strings under their distinct client-, modality-, and
epoch-specific contexts. Distinguishing $H_0$ from $H_1$ yields
$\mathcal B_2$ against HKDF pseudorandomness.

In $H_2$, each HPKE provisioning ciphertext is replaced by an
encryption of an equal-length zero string, while the simulator
preserves the corresponding honest client's internal key state.
Distinguishing $H_1$ from $H_2$ yields $\mathcal B_1$ against HPKE
confidentiality.

In $H_3$, every protected modality representation is replaced by an
encryption of an equal-length zero string. A distinguisher yields
$\mathcal B_3$ against AEAD confidentiality, while any fresh modified
ciphertext accepted by an honest client yields $\mathcal B_4$ against
AEAD integrity. This argument assumes nonce uniqueness under each
modality key.

Finally, in $H_4$, the secure-aggregation transcript is replaced by a
simulator given only
$\sum_{i\in\mathcal S_t}\alpha_i\Delta\mathbf w_i^t$.
Any distinction yields $\mathcal B_5$ against secure-aggregation
privacy. The final view depends only on permitted leakage, including
public identifiers, ciphertext lengths, protocol metadata, and the
aggregate update. Equation~(\ref{eq:security_bound}) follows by the
triangle inequality over the hybrids. \hfill$\square$

The guarantee provides upward and lateral isolation for an exposed
modality key. Compromise of the parent client key $K_i$, however,
exposes its descendant modality keys by design.

\subsection{Key Management and Overhead}
\label{subsec:key_overhead}

Table~\ref{tab:key_tradeoff} compares the proposed hierarchy with single-key and fully independent modality-key management. The exposure scope is measured under compromise of the operational encryption key.

\begin{table}[t]
\centering
\caption{Per-client key-management trade-off.}
\label{tab:key_tradeoff}
\footnotesize
\setlength{\tabcolsep}{3.2pt}
\renewcommand{\arraystretch}{1.1}
\resizebox{\columnwidth}{!}{
\begin{tabular}{lcccc}
\toprule
\textbf{Scheme}
&
\textbf{\shortstack{Persistent\\Secrets}}
&
\textbf{\shortstack{Secure\\Provisions}}
&
\textbf{\shortstack{Exposure\\Scope}}
&
\textbf{\shortstack{KDF\\Evaluations}}
\\
\midrule
Single key
&
$1$
&
$1$
&
$M_i$
&
$0$
\\
Independent modality keys
&
$M_i$
&
$M_i$
&
$1$
&
$0$
\\
\textbf{Proposed hierarchy}
&
$\mathbf{1}$
&
$\mathbf{1}$
&
$\mathbf{1}$
&
$M_i$
\\
\bottomrule
\end{tabular}
}
\end{table}

Single-key protection has low management cost but exposes all modalities after one key compromise. Independent modality keys provide fine-grained isolation but require $M_i$ long-term secrets and secure provisioning operations. Our framework provisions and stores only $K_i$, while deriving the $M_i$ modality keys locally and on demand.

Across $N$ clients, the proposed framework stores
\begin{equation}
S_{\rm ours}=N+1
\end{equation}
long-term secrets, including the root key, compared with
\begin{equation}
S_{\rm independent}
=
\sum_{i=1}^{N}M_i
\end{equation}
for independent modality-key management. It requires $N$ HPKE provisioning operations and
\begin{equation}
C_{\rm KDF}
=
\sum_{i=1}^{N}M_i
\end{equation}
local KDF evaluations per key epoch when derived keys are not cached.

Let $B_i^{(m)}$ denote the size of the protected representation for modality $m$. The client-side security overhead is
\begin{equation}
\begin{aligned}
C_{\rm sec}^{(i)}
=&\;
C_{\rm HPKE}^{(i)}
+
M_i c_{\rm KDF}
+
C_{\rm SecAgg}^{(i)}
\\
&+
\sum_{m\in\mathcal{M}_i}
\left[
C_{\rm DNA}\!\left(B_i^{(m)}\right)
+
C_{\rm AEAD}\!\left(B_i^{(m)}\right)
\right],
\end{aligned}
\label{eq:security_overhead}
\end{equation}
where HPKE is executed only during key provisioning or rotation. After authorized local decryption, the underlying gradient-based, evolutionary, or data-driven optimizer proceeds normally. Therefore, the proposed protection does not change the asymptotic complexity of local optimization; its additional cost is limited to key derivation, DNA-based transformation, symmetric encryption, and secure aggregation.

\section{Conclusion}
\label{sec:conclusion}

This paper presented a hierarchical key-managed framework for secure multimodal federated optimization. The framework distinguishes two complementary protection boundaries: modality-specific encryption protects heterogeneous local data, while secure aggregation hides individual optimization updates from the coordinator. By combining HPKE-based client-key provisioning, HKDF-based modality-key derivation, DNA-domain representation, and AES-GCM authenticated encryption, the proposed design requires only one persistent secret per client while providing cross-client and cross-modality compromise isolation. Theoretical analysis further shows that the framework composes the security guarantees of its underlying cryptographic primitives and reduces the exposure scope of an operational key compromise from all local modalities to a single modality. Future work will investigate adaptive protection policies, dynamic key rotation, and large-scale evaluation across heterogeneous multimodal optimization tasks.

%\section*{Acknowledgment}

%The preferred spelling of the word ``acknowledgment'' in America is without an ``e'' after the ``g''. Avoid the stilted expression ``one of us (R. B. G.) thanks $\ldots$''. Instead, try ``R. B. G. thanks$\ldots$''. Put sponsor acknowledgments in the unnumbered footnote on the first page.

\bibliographystyle{IEEEtran}
\bibliography{ref}

\end{document}